\documentclass[conference]{IEEEtran}
\IEEEoverridecommandlockouts
\usepackage{cite}
\usepackage{amsmath,amssymb,amsfonts}
\usepackage{algorithmic}
\usepackage{graphicx}
\usepackage{textcomp}
\usepackage{xcolor}
\usepackage{cite}
\usepackage{multirow}
\usepackage{lscape}
\usepackage{adjustbox}
\usepackage{tablefootnote}
\usepackage{tabulary}
\usepackage{array}
\usepackage{tabularray}
\usepackage{float}
\usepackage{booktabs}
\usepackage{subcaption}

\usepackage{graphicx}
\usepackage{balance}
\usepackage{expl3}
\usepackage{xcolor}
\usepackage{colortbl}
\usepackage{tabularx}
\usepackage{makecell}

\ExplSyntaxOn
\NewExpandableDocumentCommand{\colorcell}{m}
 {
  \tl_set:Nn \l_tmpa_tl { #1 }
  \regex_replace_all:nnN { [^\-\d.] } {} \l_tmpa_tl 
  \fp_compare:nTF { \l_tmpa_tl < 0 }
   { \cellcolor{red!25}#1 }
   { \cellcolor{green!25}#1 }
 }
\ExplSyntaxOff

\def\BibTeX{{\rm B\kern-.05em{\sc i\kern-.025em b}\kern-.08em
    T\kern-.1667em\lower.7ex\hbox{E}\kern-.125emX}}
\begin{document}

\title{Predicting Volleyball Season Performance Using Pre-Season Wearable Data and Machine Learning}

\author{\IEEEauthorblockN{Melik Ozolcer}
\IEEEauthorblockA{\textit{Dept. of Systems and Enterprises} \\
\textit{Stevens Institute of Technology}\\
Hoboken, NJ, USA
}
\and
\IEEEauthorblockN{Tongze Zhang}
\IEEEauthorblockA{\textit{Dept. of Systems and Enterprises} \\
\textit{Stevens Institute of Technology}\\
Hoboken, NJ, USA
}
\and
\IEEEauthorblockN{Sang Won Bae}
\IEEEauthorblockA{\textit{Dept. of Systems and Enterprises} \\
\textit{Stevens Institute of Technology}\\
Hoboken, NJ, USA
}
}

\maketitle

\begin{abstract}
Predicting performance outcomes has the potential to transform training approaches, inform coaching strategies, and \textcolor{black}{deepen} our understanding of the factors \textcolor{black}{that contribute to} athletic success. Traditional non-automated data analysis in sports \textcolor{black}{are often difficult} to scale. \textcolor{black}{To address this gap,} this study \textcolor{black}{analyzes} factors influencing \textcolor{black}{athletic} performance by leveraging passively collected sensor data from smartwatches and ecological momentary assessments (EMA). The study aims to differentiate between 14 collegiate volleyball players who go on to perform well or poorly, using data collected prior to the beginning of the season. This is achieved through an integrated feature set creation approach. The model, validated using leave-one-subject-out cross-validation, achieved promising predictive performance (F1 score = 0.75). Importantly, by utilizing data collected before the season starts, our approach offers an opportunity for players predicted to perform poorly to improve their projected outcomes through targeted interventions by virtue of daily model predictions. The findings from this study not only demonstrate the potential of machine learning in sports performance prediction but also shed light on key features along with subjective psycho-physiological states that are predictive of, or associated with, athletic success.

\end{abstract}


\section{Introduction}
Performance prediction has gained popularity in sports as a research and practical area \cite{clemente2017internal}. Understanding an athlete's performance earlier on could transform training approaches \cite{afonso2012tactical}, determine game strategies \cite{rein2016big}, and deepen understanding of factors influencing athletics and team success \cite{passfield2017mine}. Traditional methods of predicting athletic performance often involve invasive and labor-intensive data collection techniques, making it challenging to scale up these approaches, and frequently lack official match data feedback as labels \cite{claudino2019current}. 

With the emergence of artificial intelligence (AI) in sports, automating performance prediction through sensor data-driven approaches has become increasingly feasible. Understanding performance in sports enables athletes and coaches to identify strengths and weaknesses, leading to strategic adjustments in practices to enhance athlete skill levels \cite{bishop2008applied}. To maximize athletic success, especially in team sports, understanding early signals that affect performance is crucial for optimal game planning. Performance tracking can establish measurable team goals \cite{10.1145/3173574.3173745, niess2020exploring}, increasing accountability and commitment \cite{wozniak2020exploring}.

Wearable technologies have found extensive applications in personal informatics, particularly in health, sports, and well-being \cite{genaro2015overview}. Research has demonstrated the effectiveness of these self-tracking technologies \cite{kim2017omnitrack, rapp2020self} in monitoring physiological signals, biometrics, movements, and behavioral patterns, including sleep \cite{li2020extraction, fullagar2015sleep}. However, predicting performance in naturalistic environments, particularly in volleyball, remains under-explored. Existing studies have predominantly focused on exercise-specific sensor data collection using GPS sensors \cite{jung2021lax, claudino2019current} or other methods requiring human input that are difficult to scale \cite{claudino2019current}.

In this study, we introduce a novel approach using Fitbit sensor data to predict the season-long performance of collegiate volleyball players. We collect data from N=14 participants wearing Fitbit Charge 5 devices over ~26 weeks, alongside ecological momentary assessments (EMA) to understand how subjective psycho-physiological states relate to performance. Using the first 12 weeks of data (prior to season start), we train machine learning models to predict players' season-average performance. Our contributions include:

\begin{itemize}

\item Development of predictive models using official match outcome ground truths, enabling timely interventions

\item Ablation studies identifying predictive power of different season phases and feature categories

\item Statistical analyses between athletes' perceived performance reports and volleyball match metrics

\item Correlation analyses between EMA survey reports and hitting percentages to examine relationships between subjective states and recorded performance

\item Implementation of a three-step feature extraction approach capturing cardiovascular and respiratory signals beyond movement and sleep
\end{itemize}

Using leave-one-subject-out (LOSO) cross-validation, our models achieve an F1 score of 0.75, demonstrating the feasibility of predicting volleyball players' season-long performance through passive sensing data.

\section{Related Works}
\subsection{Determining Factors of Athletic Performance}
Athletes train to increase their skill, physical capacity, and consequently performance. Increases in performance levels are accomplished with gradually increased training frequency, intensity and training load \cite{american2001team}. Previous studies in the field indicate that the key factors influencing athlete performance include innate skill level, physical fitness and health, physical properties \cite{farrow2017development}, rest \cite{thun2015sleep, kellmann2018recovery}, nutrition \cite{maughan2012nutrition, brotherhood1984nutrition}, hydration \cite{murray2007hydration}, mental health and psycho-physiological states \cite{bali2015psychological, donohue2018controlled, prapavessis2000poms}, and tactical strategies \cite{mcpherson2007tactics, peters2013performance}. Studies emphasize the need for holistic approaches to understand performance, considering not just physical training but also recovery, physiological signals, and mental well-being \cite{kellmann2018recovery}.

\subsection{Health Monitoring and Health as a Prerequisite of Performance}
Prior research has examined health monitoring through collecting mobile app usage data to infer objective and subjective well-being, and mental health, \cite{hillebrand2020mobisenseus, epstein2020prediction, bauer2012can}. Researchers also evaluated prediction of subjective well-being states through wearable sensor data and EMA surveys \cite{jaques2017predicting}. Further, health monitoring has been an attractive research area in predicting hospital readmission of post-surgery patients, and sedentary behavior has been extensively linked to being predictive of worse health outcomes \cite{bae2016using, fisher2013mobility}. 
In another study, researchers argued that sensor data can be used to generate health profiles to automatically predict health status \cite{kelly2017automatic}.

As pointed out by the vast body of exercise physiology literature, the assessment of health and fitness levels must be contextualized to individuals. These assessments are achieved through continuous or periodical monitoring of physiological biomarkers \cite{lee2017biomarkers}. Common biomarkers used to assess health and fitness levels include VO$_{2\text{max}}$ \cite{bassett2000limiting}, where higher levels indicate better athletic performance and cardiovascular fitness; resting heart rate (RHR), with lower RHR signifying higher cardiovascular fitness \cite{jensen2013elevated}; heart rate variability (HRV), with higher HRV indicating better cardiovascular health \cite{plews2013evaluating}; and more intricate measurements that concern underlying long-term variations in heart rate, such as detrended fluctuation analysis (DFA), which has proven to be useful for monitoring physical responses to exercise intensities \cite{gronwald2019correlation}.

In clinical settings, there have been efforts to measure variability in heart rate using numerous approaches. DFA \cite{gronwald2019correlation, mateo2022validity, rogers2021detection, zhangpredicting, mao2012integrated}, various Entropy algorithms (Sample Entropy, Approximate Entropy \cite{pincus1991approximateentropy, richman2004sample}, second-order statistical properties (Inertia, Local Homogeneity, Energy) of heart rate have been used to detect abnormalities in heart rate and mapped to clinical outcomes \cite{zhangpredicting, mao2012integrated}, as well as being associated with biochemical markers in athletes and relationship between training and heart rate patterns \cite{lai2023use}. In this work, we extend the features that represent cardiovascular functions, extend their applications to sleep patterns and SpO${_2}$ levels, and propose their use in athletic performance prediction \cite{martin2021oxygen}.

\subsection{Performance Prediction in Sports}
Athletic performance prediction has been extensively studied \cite{feely2023modelling, claudino2019current}, particularly in sports like running \cite{keogh2019prediction, emig2020human}, cycling \cite{kholkine2020machine}, tennis \cite{panjan2010prediction}, and soccer \cite{abt2009use}. Such predictions are crucial for goal-setting and workout planning \cite{feely2023modelling}. However, research on individual athlete performance prediction in volleyball remains limited, with most existing work focusing on game tactics and team-wide analyses \cite{claudino2019current}.

The most relevant study by Leeuw et. al \cite{de2022modeling} combined jump characteristics sensor data and subjective wellness reports to model attack and defense performance separately. Their findings revealed that jump load and strength training significantly impact competition performance, achieving MAE of 0.91 for offense and 0.75 for defense on a 10-point scale. However, their approach relied on basic 10-fold CV without subject-wise splits and required manual human input, limiting its real-world scalability. Our study addresses these limitations by proposing an automated data collection method using only wearable device sensor data.

\section{Method}
This IRB-approved study \textcolor{black}{(\#2022-050)} recruited 17 participants (aged 18-22, mean=20.5) from the Men's volleyball team. Participant recruitment occurred in two phases: 10 players joined in October 2022, and 7 additional players joined in January 2023. \textcolor{black}{Athletes were asked to wear Fitbit devices continuously, including during training sessions when feasible, although actual compliance could vary.} Participants were provided Fitbit devices and setup instructions. Data collection spanned 26 weeks, with our \textcolor{black}{modeling} analysis focusing on ~12 weeks of pre-season data.

Participants also completed daily ecological momentary assessment (EMA) surveys at 10AM and 9PM, rating their perceived injury risk, readiness, recovery, soreness, tiredness, mood, stress, sleep quality, performance, and productivity on a 1-7 scale. Weekly compensation (\$25/\$20/\$15) was provided based on compliance rates, and participants could withdraw at any time.

\subsection{Timeline of the Volleyball Season}
We categorized the volleyball season into four phases in the calendar year to understand the impact of the sensor data collected during different times of the year. Because different times in the season has different goals, intensity, training regimen and expectations according to the team coach. 

The volleyball season takes place in the Spring semester, but players train individually throughout the year. The team conducts 15 practice sessions over a span of 26 days in the Fall practice season (\textbf{Phase 1}). During this phase, the team integrates newly joined players, goes through tactical drills, and plays friendly matches with other schools to prepare for the upcoming season. As the end of the semester nears, team practices are put on hold until after the new year. We label the end of the semester and the winter break period as \textbf{Phase 2}.

\begin{table}[t]
\centering
\caption{Timeline of the Volleyball Season}
\label{tab:volleyball-timeline}
\begin{tabular}{llr}
\toprule
\textbf{Phase} & \textbf{Period} & \textbf{Duration} \\
\midrule
Phase 1 & Fall Practice Season & Oct 23--Nov 18 \\
Phase 2 & End of Semester + Winter Break & Nov 18--Jan 3 \\
Phase 3 & January Practice Block & Jan 3--Jan 13 \\
Phase 4 & NCAA Volleyball Season & Jan 13--Apr 28 \\
\bottomrule
\end{tabular}
\end{table}

Then, the team holds a January practice camp where they get together after the break and have a high intensity training camp for 10 days, to prepare for the NCAA season. We labeled this period as \textbf{Phase 3}. At the end of the second week of January, the team plays their first official match to kick off the 2023 NCAA men's volleyball season that lasts until the end of April (Phase 4). We statistically analyze and build predictive models with data from different parts of the year to examine the predictive power of body signals in different phases. 

\subsection{Data Preparation and Feature Extraction}
Time series data including step count, distance, calories (1-min granularity), heart rate (~8.5/min granularity), daily HRV, RHR, sleep data, breathing rate, VO$_{2\text{max}}$, and SpO$_{2}$ (1-min granular) were collected from the Fitbit API. For time series data, we computed various statistical aggregations to convert them into daily granularity.

\subsubsection{Movement} 
Movement features include daily aggregations of step count, distance, and calorie expenditure. We extracted sedentary metrics following \cite{bae2016using}, including sedentary time (categorizing step counts into three levels: 0-33, 34-67, 68+), sedentary bouts (zero-step time frames), and sedentary breaks (post-sedentary bout activity levels).

\subsubsection{Sleep} 
Sleep features were derived from sleep stage times (deep, light, REM, wake). We computed Detrended Fluctuation Analysis (DFA) from sleep stage time series data, resampled to 1-minute intervals, using 10-60 minute windows for each sleep event \cite{zhang2022predicting}.

\subsubsection{Cardiovascular} 
For heart rate data, we extracted statistical aggregations, skewness, kurtosis, DFA, second-order statistics, and entropy \cite{gronwald2019correlation, mateo2022validity}. Second-order statistical features included inertia, local homogeneity, correlations, and energy \cite{mao2012integrated}. Day-over-day changes in RHR and HRV were also computed as literature suggests HRV changes can indicate changes in exercise regimen \cite{makivic2013heart}.

\subsubsection{Respiratory}
Respiratory features include daily BR and VO$_{2\text{max}}$ values, along with statistical aggregations of SpO$_{2}$ data. From SpO$_{2}$ data, we computed basic statistics, skewness, kurtosis, DFA in 10-60 minute windows, and estimated the Hurst Exponent.

\subsubsection{High-Level Features}
DFA was applied to heart rate, sleep stage, and SpO$_{2}$ data, as it is suitable for non-stationary time series data \cite{hu2001effect}. The Hurst Exponent was estimated using the slope of the log-log plot of DFA values across varying window sizes.

\subsection{Ground Truths}
\begin{figure*}[t]
\centering
\includegraphics[width=0.225\textwidth]{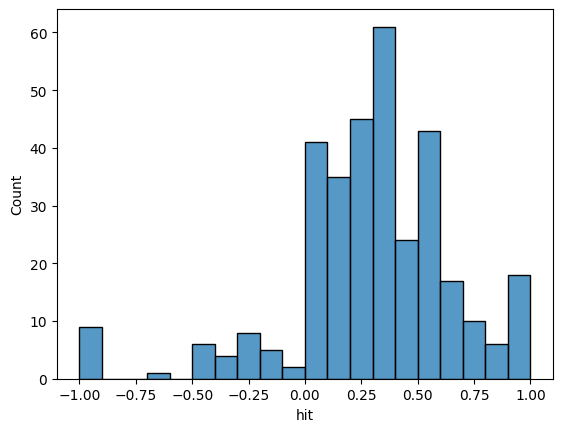}
\hspace{0.5mm}
\includegraphics[width=0.25\textwidth]{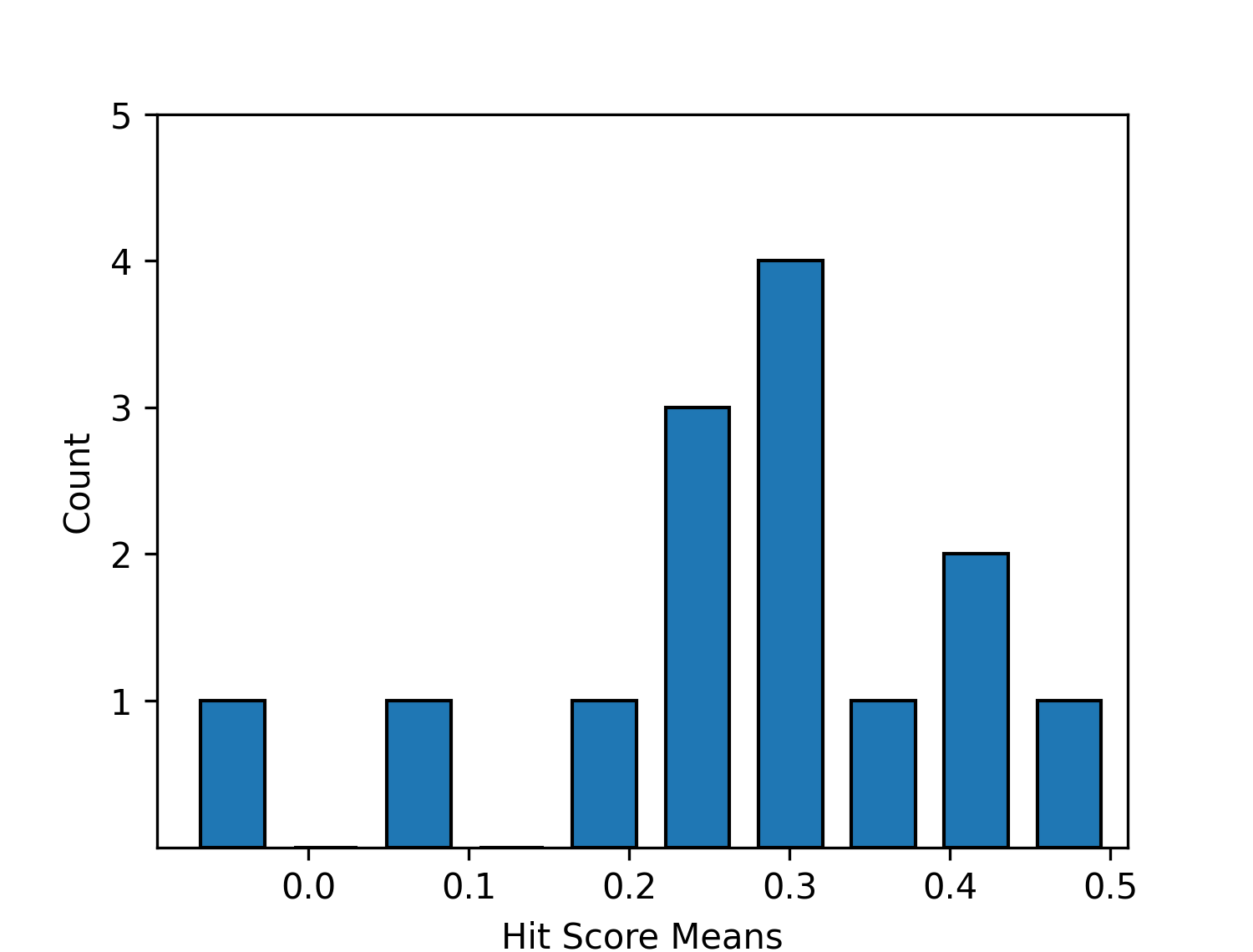} 
\hspace{0.5mm}
\includegraphics[width=0.25\textwidth]{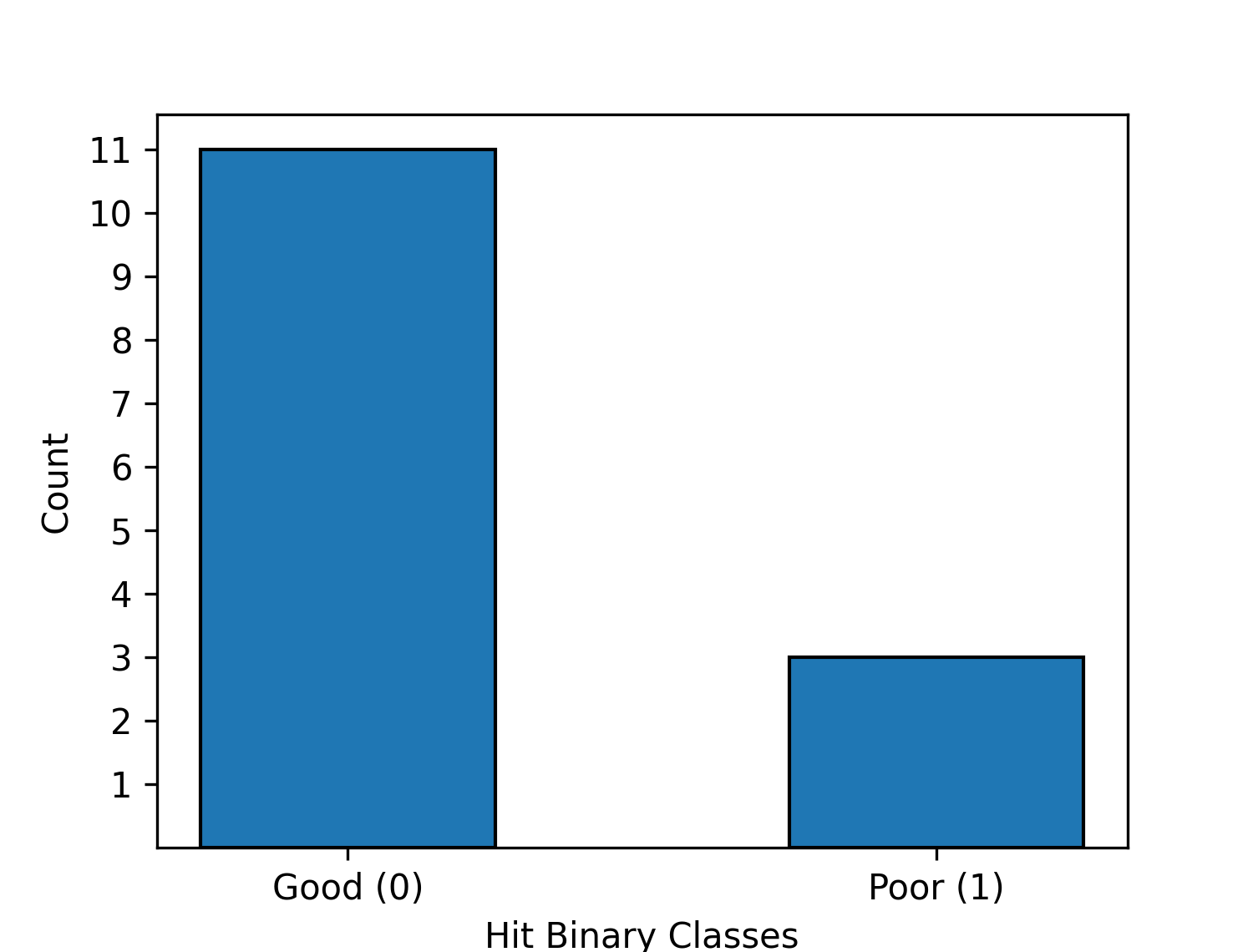} 
\caption{Data Distributions of Hit scores throughout the Season (left), Season Hit Averages for Athletes (middle), Class Distributions (right). \textcolor{black}{The season-long average hit percentage scores of athletes were computed (collected through Phase 4), then binarized. The binary performance outcomes were then used to label the sensor data in Phases 1,2,3.}}
\label{fig:distributions}
\end{figure*}

With the guidance of the coaches of the men's volleyball team, we chose the hit percentage \cite{sanders2018accelerometer, DartmouthVolleyball2022}, a critical measure of attacking efficiency, as our quantification of athlete performance. The hit percentage in volleyball is calculated as \(\text{Hit \%} = \frac{\text{Kills} - \text{Errors}}{\text{Attempts}}\), where \textit{Kills} represent successful attacks that result in earned point, \textit{Errors} are offensive attacks that result in opposition team earning a point, and \textit{Attempts} are the total attack attempts. This metric provides a quantitative assessment of a player's offensive effectiveness, and overall game performance.
We fetched the game statistics data for each player from the official website of the team where the box scores for each official match are logged. Then, we computed the average season hit percentage of each player and labeled the sensor data with their corresponding hit percentage average. We categorized the hit percentage scores as "good" and "poor", based on the criteria established by the head coach of the men's volleyball team. We categorized hit scores above 0.2 as "good" (class 0), and those below 0.2 as "poor" (class 1). The data distributions for additional insights about the ground truth is presented in Figure \ref{fig:distributions}. \textcolor{black}{While player performance is multidimensional, the 0.2 cutoff directly reflects the expert coach’s threshold for classifying good versus poor performance. We acknowledge that more fine-grained or regression-based models could provide additional insights, but this binary classification was a clear, expert-driven approach for our initial study.}


\subsection{Rationalization of Hit Percentage as a Ground Truth}

\textcolor{black}{Table \ref{tab:perceived_performance_correlations} presents the correlations between various volleyball game stats metrics and the perceived performance of players, as reported later in the day. These reports were recorded in the night survey, and the players reported them retrospectively for the match played earlier in the same day. The game metrics presented are taken directly from the NCAA website box scores, and these are all the officially recorded volleyball game metrics by the NCAA.}

The results show that the \textit{Hit Percentage} has the strongest positive correlation with perceived performance ($\rho$ = 0.346, p = 0.002). This finding suggests that players with higher hitting percentages tend to have higher perceived performance ratings. Interestingly, the number of \textit{Points} scored does not exhibit a significant correlation with perceived performance ($\rho$ = 0.066, p = 0.572), despite being thought of as a valuable game metric.

Among the other metrics, \textit{Reception Errors} show a significant negative correlation with perceived performance ($\rho$ = -0.288, p = 0.012), indicating that players with higher numbers of reception errors tend to have lower perceived performance ratings. This highlights the importance of reliable ball reception in volleyball.

\begin{table}[t]
\centering
\caption{\textbf{Perceived Performance} Correlations with Volleyball Game Metrics. }
\begin{tabular}{lrr}
\toprule
 & \textbf{p-value} & \textbf{Spearman's \( \rho \)} \\
\midrule
Kills & 0.706 & 0.044 \\
Digs & 0.684 & 0.047 \\
Service Aces & 0.610 & 0.059 \\
Points & 0.572 & 0.066 \\
Block Solos & 0.523 & -0.074 \\
Total Attempts & 0.480 & -0.082 \\
Service Errors & 0.321 & 0.115 \\
Block Errors & 0.300 & -0.120 \\
Ball Handling Errors & 0.298 & 0.121 \\
Assists & 0.186 & 0.153 \\
Errors & 0.074 & -0.206 \\
Reception Errors & 0.012 & \colorcell{-0.288} \\
Hit Percentage & 0.002 & \colorcell{0.346} \\
\bottomrule
\end{tabular}
\label{tab:perceived_performance_correlations}
\end{table}

The hitting percentage was chosen as the ground-truth, confirmed by the expert, the volleyball team coach, and our analysis, in which the \textit{Hit Percentage} showed the highest correlation among other game metrics with the subjective feeling of a player's performance on that day, further supporting the use of hitting percentage as a key indicator of volleyball performance.

\section{Model Development}
We train models for each of the 7 dataset phase \textcolor{black}{combinations}, using five different machine learning algorithms. We apply the same feature selection, and conduct an equal level of hyperparameter tuning for all trained models. Considering the moderate size of our study, we select relatively simple machine learning models to avoid overfitting. We evaluate the performance of our models using leave-one-subject-out (LOSO) cross-validation. 

\subsection{Feature Selection}
We employ feature selection and mitigate multicollinearity to help prevent overfitting and maintain the model's interpretability. For the data in each phase \textcolor{black}{combination}, we first check the pairwise correlation coefficients of input features. If any feature exhibits a Pearson correlation coefficient of greater than 0.7 with another feature \cite{dormann2013collinearity}, we eliminate that feature from our analysis. Next, we employ univariate feature selection by applying F-test for each of our continuous variables to test if the feature has different means between the good performing and poor performing athletes. The resulting features are kept for the model training.

\subsection{Predictive Modeling}
We deployed five different classifiers, namely, XGBoost \cite{chen2016xgboost} (XGB), LightGBM \cite{ke2017lightgbm} (LGBM), random forest classifier \cite{breiman2001random} (RFC), linear support vector machine (SVM), gaussian naive bayes classifier (GNB) on the entire sensor data (Phase 1+2+3) to determine the best performing classifier for this task. Then with the selected machine learning model, we compare the predictive performance that we get from sensor data collected during different phases \textcolor{black}{and phase combinations}.

\subsubsection{Preprocessing} 

All trained models followed a uniform data preprocessing approach. For every trained model, we first impute missing values with column means. Then, we normalize our data between 0 and 1. We sample the data distribution strictly from the training data, to prevent any information leakage that would not occur in a real world scenario.
We then incorporate the Synthetic Minority Over-sampling Technique (SMOTE) in processing our training data. SMOTE generates synthetic minority samples, helping to ensure our model adequately represents all classes \cite{chawla2002smote}.

\subsection{Model Evaluation}
We employed LOSO cross-validation, where data from one athlete is reserved for evaluation while training on all others, simulating real-world prediction scenarios. To ensure robustness and reduce overfitting, we bootstrapped the LOSO process with 10 iterations. Model performance was evaluated using accuracy, F1-score, recall, precision, AUROC, and AUPRC, with reported values averaged across iterations (standard deviations in parentheses).

\begin{figure*}[!t]
    \centering
    \includegraphics[width=0.75\textwidth]{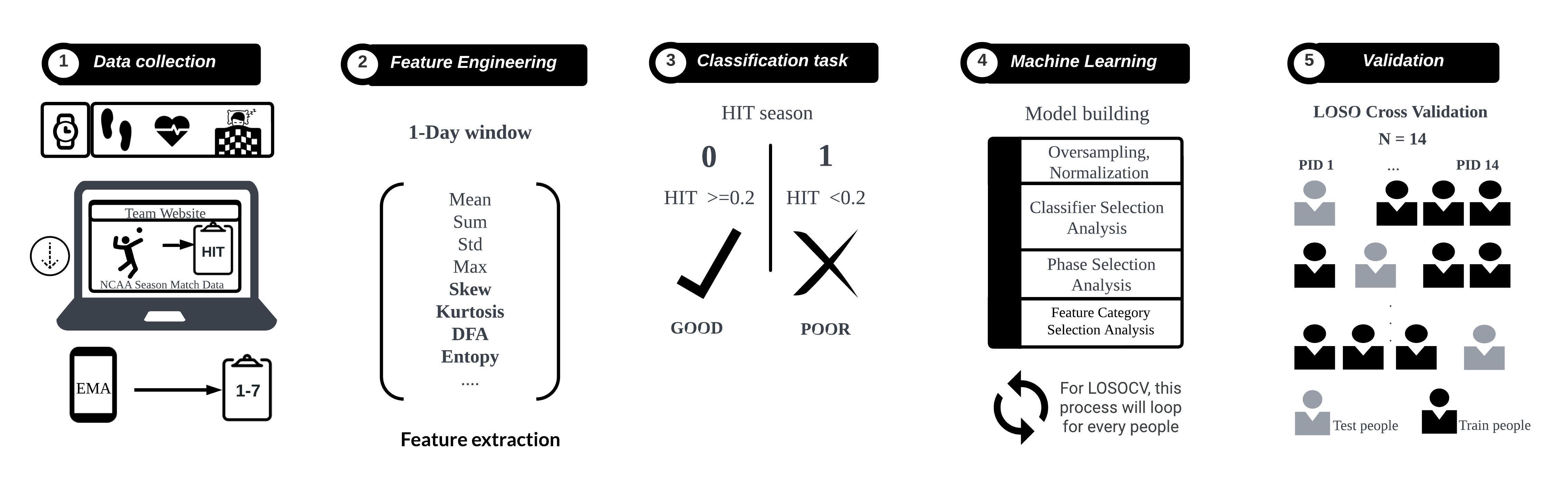}
    \caption{Overview of Machine Learning Pipeline.}
    \label{fig:sport_pl}
\end{figure*}

\section{Exploratory Study in Volleyball Performance}
Out of the 17 total participants, three were excluded from all analyses: one due to playing an insufficient number of games (only 1 match), another because their position (libero) does not involve a tracked hit score, and the third for providing inadequate amount of data (only 1 day). Consequently, the models included a maximum of 14 players and a minimum of 8 players, depending on the specified phase.

Among the 14 players, 11 had a season hit average greater than 0.2 (good), while 3 had an average below 0.2 (poor). 1/3 poor performers enrolled in our study during Phase 3, while the other two enrolled in Phase 1. Notably, all 3 players in the team with season hit averages below 0.2 were included in our study (Noting that the team went on to win the national championship at the end of the season).

In our study, we are analyzing data only up to the point just before the season begins \textcolor{black}{for machine learning modeling}, focusing on future predictive implications (Data collected through \textbf{Phase 1+2+3}, ground truths collected during \textbf{Phase 4}). \textcolor{black}{We also provide statistical analyses on the data collected during Phase 4.}

\subsection{Missing Data}
\subsubsection{Sensor Data}
\begin{figure*}[t]
   \begin{subfigure}[b]{0.23\textwidth}
        \centering
        \includegraphics[width=\textwidth]{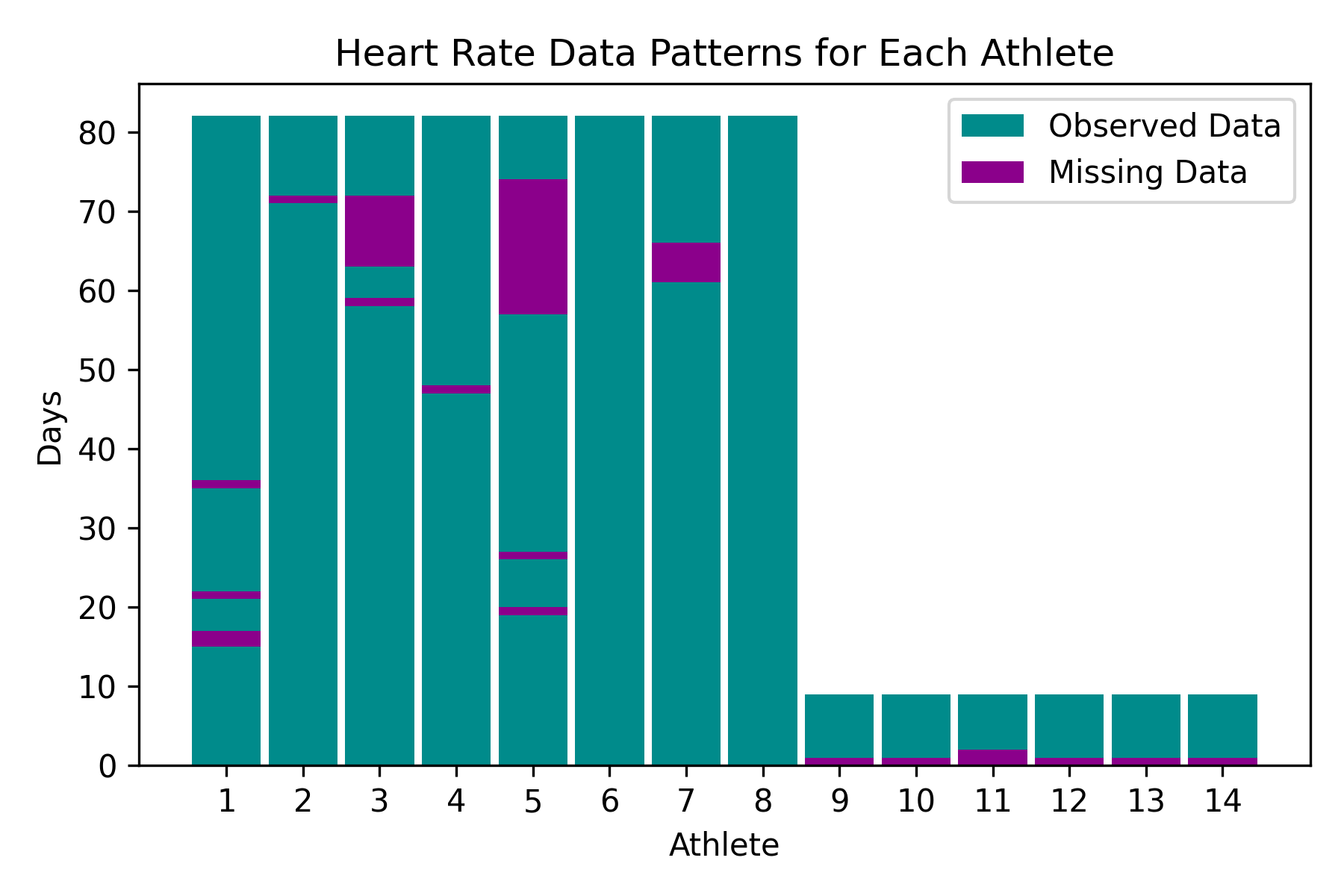}
        \caption{Heart rate data missingness patterns for each athlete (days with $>$ 0 HR records).}
   \end{subfigure}
   \hfill
   \begin{subfigure}[b]{0.23\textwidth}
        \centering
        \includegraphics[width=\textwidth]{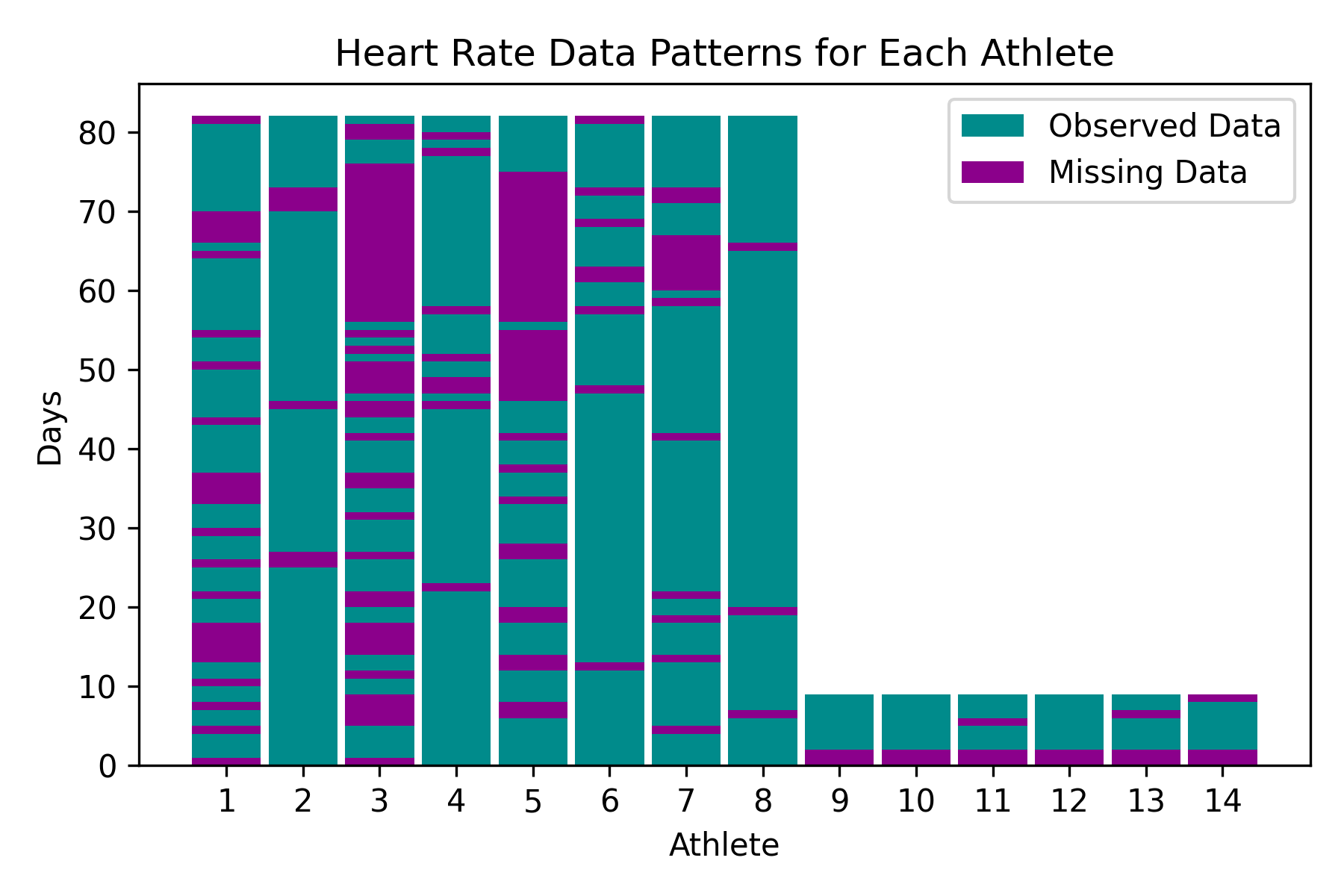}
        \caption{Heart rate data missingness patterns after 70\% daily compliance exclusion.}
   \end{subfigure}
   \hfill
   \begin{subfigure}[b]{0.23\textwidth}
        \centering
        \includegraphics[width=\textwidth]{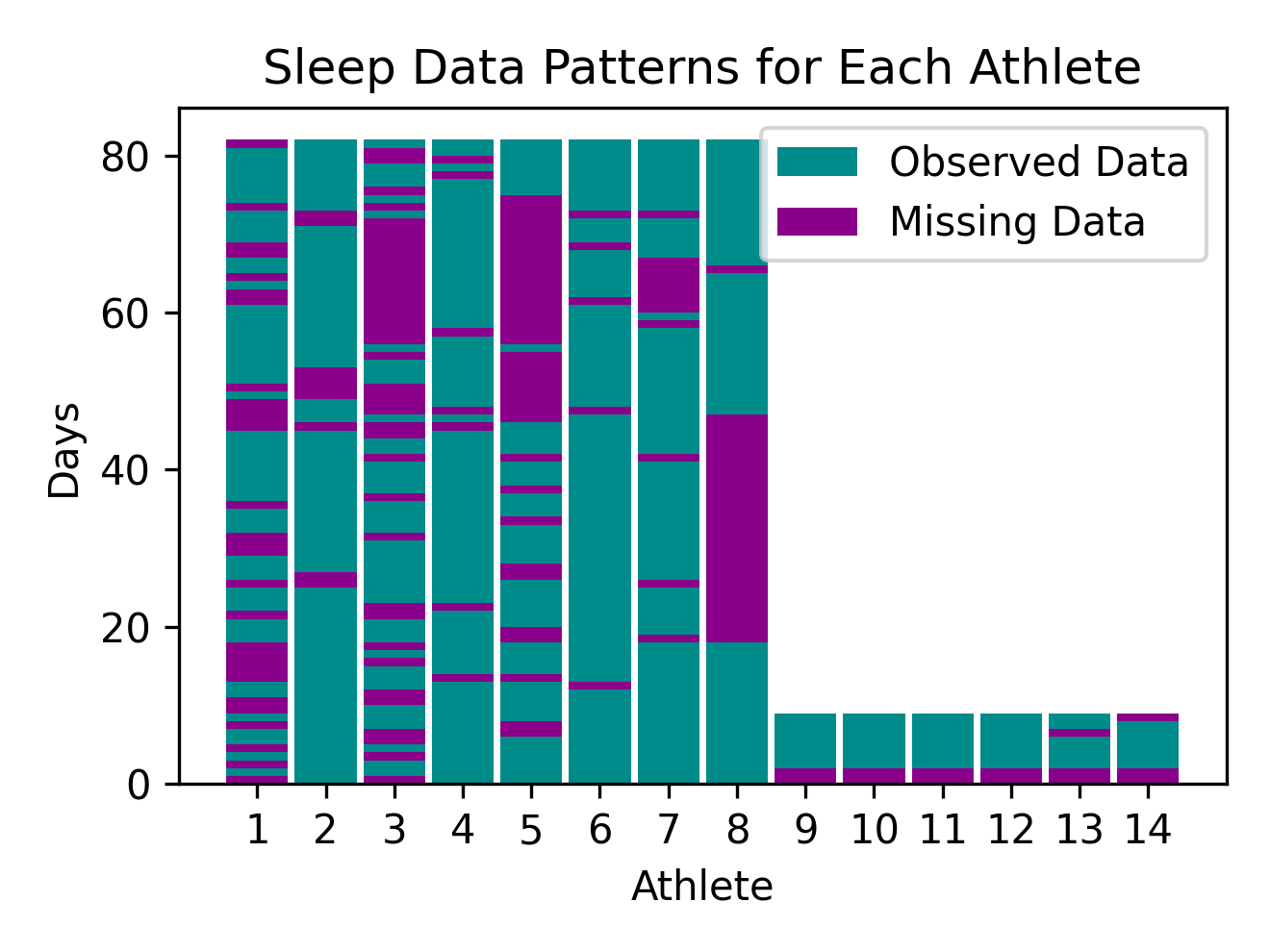}
        \caption{Sleep data missingness patterns for each athlete.}
        \label{fig:missing_sleep}
   \end{subfigure}
   \hfill
   \begin{subfigure}[b]{0.23\textwidth}
        \centering
        \includegraphics[width=\textwidth]{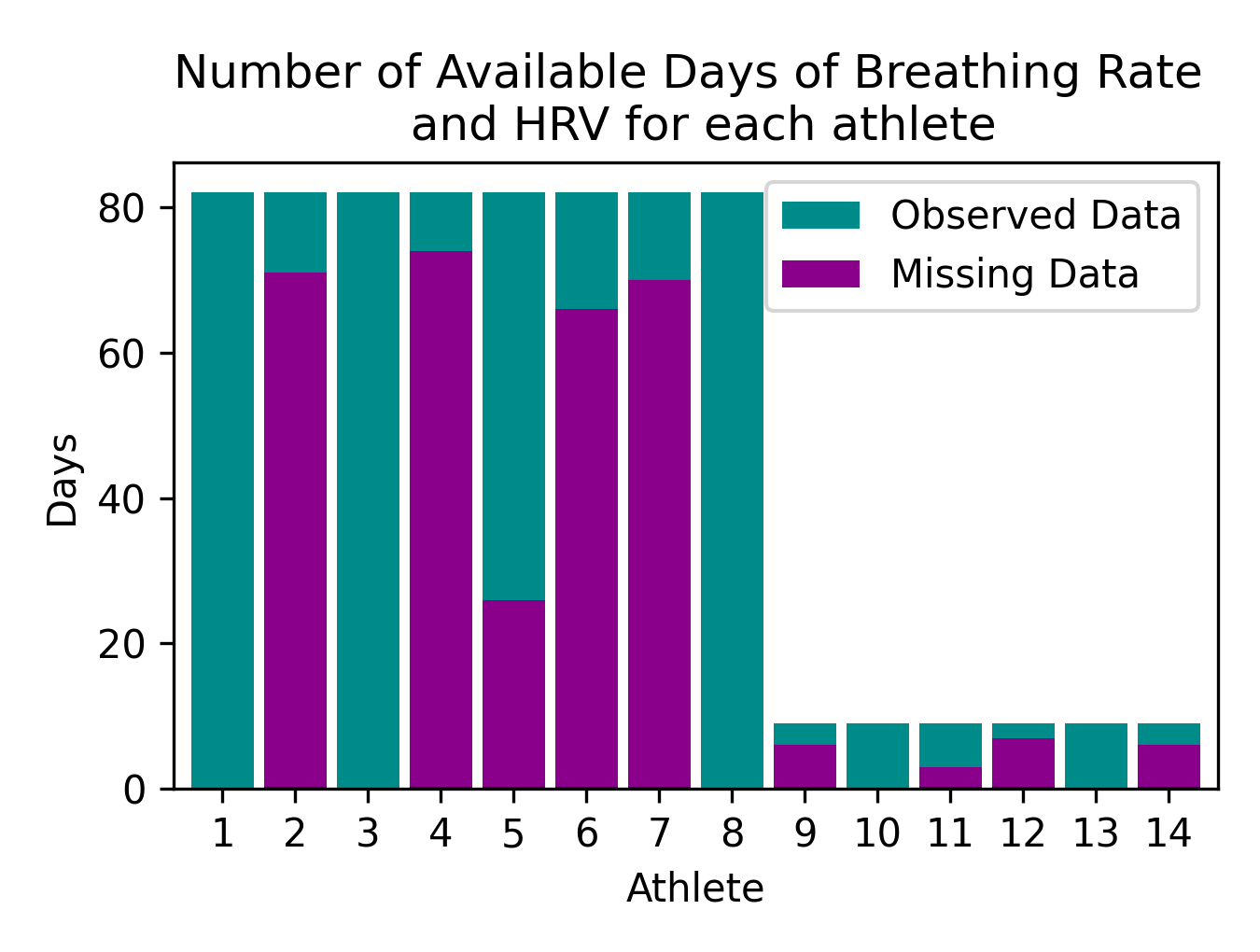}
        \caption{Breathing rate and heart rate variability (HRV) data.}
        \label{fig:missing_hrv}
   \end{subfigure}
\caption{Missing data visualization for heart rate, sleep, and available days of breathing rate and heart rate variability data per participant during Phase 1+2+3. (a--c) show missing data patterns; (d) shows number of available days per feature.} 
\label{fig:missing_hr}
\end{figure*}

\textcolor{black}{During Phases 1+2+3}, the average length of available wearable data was 47.36 days, with a range from 7 to 82 days, totaling 663 days. We established a strict daily compliance rate cutoff of 70\% to ensure a high level of data quality. We found that Fitbit devices record heart rate readings at an average rate of 8.5 times per minute (std $\cong$ 2). Therefore, we excluded days where the number of heart rate readings was less than 8,768 (0.7 x 1440 x 8.5). Following this, the average length of wearable data available was reduced to 38.86 days, ranging from 6 to 79 days, totaling 544 days. The patterns of missingness and the change in available data before and after the exclusion is shown in Figure \ref{fig:missing_hr}.

In the resulting dataset, missing data was still prevalent due to various reasons. Occasionally, sleep data records were lost when participants did not synchronize their Fitbit devices for extended periods or did not wear their device (see Figure \ref{fig:missing_sleep}). Non-recorded sleep resulted in the absence of feature categories that are computed by Fitbit during the sleep, such as RHR, VO$_{2\text{max}}$, HRV, breathing rate, and SpO${_2}$. Additionally, HRV and breathing rate data were not collected from 5 participants who opted out of the collection of these data categories, as shown in Figure \ref{fig:missing_hrv}. There were occasional random missingness of certain features due to unsatisfied conditions in the computation of features. Overall, the dataset had 6263 missing values across 544 data points and 72 features, amounting to ~16\% of all data points.

\subsubsection{EMA Survey Responses}
Participants received notification emails to the surveys every morning and night, also had the option to report in the afternoons as well. For our 1-day granular data, we calculated the daily average of all responses per participant. The average number of available days with a survey data \textcolor{black}{during Phase 1+2+3} was 27.86 days, ranging from 1 day to 75 days, totaling 390 days.

\subsection{Statistical Analysis of EMA Surveys and Performance}
In this section, we examine the relationships between various psycho-physiological factors captured through EMA surveys, and the athletes' season average hit scores in an available-case-analysis. To achieve this, we employed Spearman's rank correlations \cite{gauthier2001detecting}, suitable for the ordinal and non-parametric nature of our survey answers. Table \ref{table:correlation} presents the Spearman's rank correlation coefficients and respective p-values for the EMA survey items that were found have statistically significant correlations with the season hit average (p-value $<$ 0.05) in each phase. We see that perceived stress, perceived injury risk and perceived productivity are prevalent in most phases, stress being in 6 out of 7, injury risk and productivity being in 4 of the 7 phases. The strength of correlations slightly varied across phases, however the directions remain unchanged. Additionally, we see that perceived recovery (Spearman's \( \rho = 0.178\), \( p = 0.039\)) had a positive correlation with performance during the fall practice season (Phase 1). 

Over the course of the fall practice season, winter break and January practice camp (Phase 1+2+3), we observe that perceived injury risk (\( \rho = 0.116\), \( p = 0.022\)) and perceived productivity (\( \rho = 0.140\), \( p = 0.006\)) were positively correlated with average hit scores. Productivity (which encompasses their school and personal lives) is a trait of conscientious individuals, and the relationship in perceived productivity may suggest the existence of a relationship between conscientiousness and motivated personalities and athletic performance, as highlighted by the existing literature \cite{mirzaei2013relationship, tedesqui2018comparing}. Conversely, perceived  stress showed a more significant and a negative correlation with hit performance (\( \rho = -0.228\), \( p < 0.001\)), highlighting the negative impact of stress on athletic success \cite{brandt2017perceived, fullagar2015sleep}.

{
\renewcommand{\arraystretch}{0.85}
\begin{table}
    \centering
    \caption{Spearman's rank correlation coefficients for the EMA surveys and season average hit performances across phases.}
    \label{table:correlation}
    \setlength{\tabcolsep}{3pt}  
    \small  
    \begin{tabular}{llrr}
    \toprule
    \textbf{Phase} & \textbf{EMA Survey Item} & \textbf{p-value} & \textbf{Spearman's \(\rho\)} \\
    \midrule
    \multirow{3}{*}{Phase 1} 
        & Perceived Recovery & 0.039 & 0.178 \\
        & Perceived Productivity & 0.004 & 0.247 \\
        & Perceived Stress & $<$0.001 & -0.352 \\
    \midrule
    \multirow{2}{*}{Phase 2}
        & Perceived Injury Risk & 0.012 & 0.182 \\
        & Perceived Stress & $<$0.001 & -0.265 \\
    \midrule
    \multirow{2}{*}{Phase 3}
        & Perceived Injury Risk & 0.033 & 0.263 \\
        & Perceived Productivity & 0.021 & 0.283 \\
    \midrule
    Phase 1+2 & Perceived Stress & $<$0.001 & -0.297 \\
    \midrule
    \multirow{2}{*}{Phase 1+3}
        & Perceived Stress & 0.010 & -0.181 \\
        & Perceived Productivity & $<$0.001 & 0.243 \\
    \midrule
    \multirow{2}{*}{Phase 2+3}
        & Perceived Stress & 0.002 & -0.190 \\
        & Perceived Injury Risk & $<$0.001 & 0.206 \\
    \midrule
    \multirow{3}{*}{Phase 1+2+3}
        & Perceived Injury Risk & 0.022 & 0.116 \\
        & Perceived Productivity & 0.006 & 0.140 \\
        & Perceived Stress & $<$0.001 & -0.228 \\
    \bottomrule
    \end{tabular}
\end{table}
}

The surprising positive relationship between the injury risk and hit scores demands further investigation, because it likely reflects the existence of nuanced relationships between various factors. Potential explanations for this observation could be (a) a reporting bias where the players who generally perform better are more likely to have an increased sense of attention to their physical state; (b) better risk mitigation by players who perceive higher injury risk and deciding to be more cautious during the season, avoiding reckless actions that could lead to actual injuries and ultimately leading to better athletic performance; (c) increased motivation for conditioning by athletes who perceive higher risk levels of injury , enhancing their performance; or (d) it could be a sign of more intensive training, and higher levels of accumulated strain during the off-season, indicating better preparation for the upcoming season.

These results emphasize the importance of psycho-physiological evaluations in predicting athletic success, complementing analyses of physiological and behavioral patterns collected through wearable devices.

\subsection{Evaluation of Predictive Models}
In this subsection, we discuss the predictive model selection and the choice of season phase. First, we select the machine learning classifier, and with the selected classifier, we select the best phase. 

\subsubsection{Classifier Selection}
Different models XGBoost, LightGBM, RFC, gaussian NB, and linear SVM were trained on the \textbf{Phase 1+2+3} sensor data. We conducted 10 simulations for each model, with the same set of features and roughly equally rigorous optimization processes through Optuna, presented in Table \ref{tab:classifier_comp}. We used t-tests to test the hypothesis that the differences between the average F1-scores of the best performing classifier and the other classifiers are statistically significant. In our experiment, XGBoost outperformed other machine learning models. For simplicity, we only report the results from the best performer (XGB) and the second best performer (RFC) (p-value = 0.027). Therefore, we select XGBoost as the classifier of our choice for further analyses. 

\begin{table*}[t]
    \centering
    \caption{Predictive Performance of Different Machine Learning Models on the \textbf{Phase 1+2+3} Data.}
    \label{tab:classifier_comp}
    \begin{tabular}{lllllll}
    \toprule
    Classifier & Accuracy & F1-score & Precision & Recall & AUROC & AUPRC \\
    \midrule
    GNB & 0.5790 (0.000)  & 0.3324 (0.000)& 0.3065 (0.000)& 0.3631 (0.000)& 0.4886 (0.000)& 0.2836 (0.000) \\
    LGB & 0.7884 (0.008)& 0.5605 (0.022)& 0.6998 (0.021)& 0.4682 (0.028)& 0.6774 (0.020)& 0.5784 (0.025)\\
    RFC & 0.7949 (0.007) & 0.5681 (0.014)& 0.7243 (0.022)& 0.4675 (0.015)& 0.6693 (0.018)& 0.5986 (0.021)\\
    SVM & 0.5422 (0.019) & 0.4123 (0.020)  & 0.3276 (0.017)  & 0.5562 (0.026) & 0.5795 (0.016)  & 0.4371 (0.028) \\
    \textbf{XGB} & \textbf{0.7925 (0.004)} & \textbf{0.5819 (0.011)}& \textbf{0.6949 (0.008)}& \textbf{0.5007 (0.015)}& \textbf{0.6807 (0.013)}& \textbf{0.6030 (0.013)}\\
    \bottomrule
    \end{tabular}
\end{table*}

\subsubsection{Phase Selection}
In Table \ref{tab:model_performance}, we present the predictive performances achieved in each phase for the selected classifier, XGBoost. The best individual overall performance was achieved with Phase 2+3 sensor data (F1-score = 0.7549, AUPRC= 0.7776). Surprisingly, more samples and a bigger dataset did not result in increased predictive performance, as Phase 1+2+3 results indicate. This could be an indication of time sensitive data patterns and the changing distinguishability of classes over different phases, or a sign of overfitting in the Phase 2+3 data. Further, as expected, Phase 3 data, which spans over 9 days, resulted in the lowest predictive performance. Regardless, we select Phase 2+3 as the phase to further analyze.

\begin{table*}[t]
\centering
\caption{Predictive Performance of Different Machine Learning Models on Sensor Data Collected in Different Phases of the Season.}
\label{tab:model_performance}
\begin{tabular}{lllllllll}
\toprule
\textbf{Phase} & \textbf{Days} & \textbf{Classifier} & \textbf{Accuracy} & \textbf{F1-score} & \textbf{Precision} & \textbf{Recall} & \textbf{AUROC} & \textbf{AUPRC} \\
\midrule
Phase 1 & 26 &XGB & 0.7429 & 0.5714 & 0.5556 & 0.5882 & 0.6581 & 0.6070 \\
Phase 2 & 46 &XGB & 0.7481 & 0.5988 & 0.5882 & 0.6098 & 0.8456 & 0.6487 \\
Phase 3 & 9 &XGB & 0.6699 & 0.2273 & 0.2500 & 0.2083 & 0.6883 & 0.3579 \\
Phase 1+2 & 72 &XGB & 0.7234 & 0.5643 & 0.5374 & 0.5940 & 0.7052 & 0.5982 \\
Phase 1+3 & 35 &XGB & 0.8107 & 0.6436 & 0.7045 & 0.5924 & 0.7398 & 0.6572 \\
\textbf{Phase 2+3} & \textbf{55} &\textbf{XGB} & \textbf{0.8645} & \textbf{0.7549} & \textbf{0.7857} & \textbf{0.7264} & \textbf{0.8236} & \textbf{0.7776} \\
Phase 1+2+3 & 81 &XGB & 0.7996 & 0.5948 & 0.7143 & 0.5095 & 0.6834 & 0.6216 \\
\bottomrule
\end{tabular}
\end{table*}

\subsection{Key Features Identified Through F-Test}
In this subsection, we analyze the features that resulted from the F-test, for our most predictive dataset, Phase 2+3. Our aim is to further understand which body signals are most influential in determining the hit percentage classification. 

\begin{table*}[t]
\centering
\caption{Feature Selection Results and Group Statistics
| Note: All features shown have p $<$ 0.05. Features are sorted by F-value.}
\label{tab:feature-selection}
\begin{tabular}{lcccc}
\toprule
Feature & F-value & $p$-value & \multicolumn{2}{c}{Mean (SD)} \\ 
\cmidrule(lr){4-5}
& & & Good & Poor \\ 
\midrule
HRV & 86.279 & $<$0.001 & 49.326 (19.131) & 87.000 (29.455) \\
Breathing Rate & 60.780 & $<$0.001 & 13.985 (1.365) & 16.717 (2.102) \\
Total Sedentary Time  & 46.942 & $<$0.001 & 2033.110 (407.725) & 2351.821 (395.723) \\
HRV Change & 46.905 & $<$0.001 & -0.337 (15.862) & -0.414 (39.957) \\
HR Skewness & 19.374 & $<$0.001 & 0.939 (0.494) & 0.690 (0.486) \\
Sleep Efficiency & 15.211 & $<$0.001 & 82.624 (20.682) & 92.471 (4.509) \\
$\text{VO}_2\text{max}$ & 11.610 & 0.001 & 48.182 (2.056) & 48.878 (0.693) \\
Sedentary Break Total & 10.908 & 0.001 & 95.019 (25.822) & 104.349 (21.062) \\
Sedentary Break Std. & 10.447 & 0.001 & 0.384 (0.088) & 0.351 (0.086) \\
Heart Rate Min. & 7.173 & 0.008 & 46.536 (4.621) & 45.208 (3.419) \\
Sedentary Bout Std. & 5.375 & 0.021 & 0.481 (0.023) & 0.474 (0.032) \\
$\text{SpO}_2$ Skewness & 4.398 & 0.037 & -1.327 (1.126) & -1.631 (1.086) \\
$\text{SpO}_2$ Hurst Exp. & 4.169 & 0.042 & 1.241 (0.185) & 1.198 (0.172) \\
\bottomrule
\end{tabular}
\end{table*}

The F-test analysis identified 13 significant features ($p$ $<$ 0.05) that distinguish between good and poor volleyball performers based on their season average hit scores. The most discriminative features were related to heart rate variability (HRV) and respiratory metrics, with HRV showing the highest F-value (86.279, $p$ $<$ 0.001) where poor performers showed notably higher values (87.000 ± 29.455) compared to good performers (49.326 ± 19.131). Breathing rate was the second most significant feature (F = 60.780, $p$ $<$ 0.001), with poor performers showing higher rates (16.717 ± 2.102) than good performers (13.985 ± 1.365). Activity-related features also proved important, particularly total sedentary time and sedentary behavior patterns, where poor performers generally showed higher sedentary time (2351.821 ± 395.723 vs 2033.110 ± 407.725) and different break patterns. Sleep efficiency emerged as another significant discriminator (F = 15.211, $p$ $<$ 0.001), with poor performers showing higher efficiency (92.471 ± 4.509 vs 82.624 ± 20.682). The analysis also identified several other physiological markers including minimum heart rate, SpO2 skewness, and the SpO2 Hurst exponent, which characterizes the long-range temporal correlations in the SpO2 time series across different time scales (10-60 minutes), as significant features. \textcolor{black}{While these analyses do not directly dissect specific technical skill execution (e.g., attacking form or jump mechanics), the relationships observed here imply that sensor-derived features can capture relevant signals impacting performance outcomes.}

Interestingly, these results reveal a counterintuitive pattern where poor performers demonstrated what traditionally would be considered "better" physiological markers (higher HRV, sleep efficiency, and slightly higher VO2 max). This unexpected finding might suggest that good performers maintain a more consistent training load that results in sustained physiological stress, while poor performers may be under-training, leading to better recovery metrics but suboptimal performance outcomes. 

\subsection{Relationships between Objectively Measured Athletes' Hit Scores and Subjectively Reported Perceived Experiences through EMA Reports in Phase 4}
\textcolor{black}{The results show that the factor with the strongest positive correlation to hit scores is subjectively reported perceived performance ($\rho$ = 0.346, $p$ = 0.002). This suggests that players who reported higher levels of performance tended to have higher hit scores during Phase 4. Secondly, perceived soreness had a positive correlation with the hit score ($\rho$ = 0.169, $p$ = 0.045)}

\begin{table}[t]
\centering
\begin{tabular}{lrr}
\toprule
 & \textbf{p-value} & \textbf{Spearman's \( \rho \)} \\
\midrule
Perceived Soreness & 0.045 & \colorcell{0.169} \\
Perceived Performance & 0.002 & \colorcell{0.346} \\
\bottomrule
\end{tabular}
\caption{Daily Hit correlations in Phase 4}
\label{tab:daily_hit_correlations_phase4}
\end{table}

\begin{figure}[t]
    \centering
    \begin{subfigure}[b]{0.48\columnwidth}
        \centering
        \includegraphics[width=\textwidth]{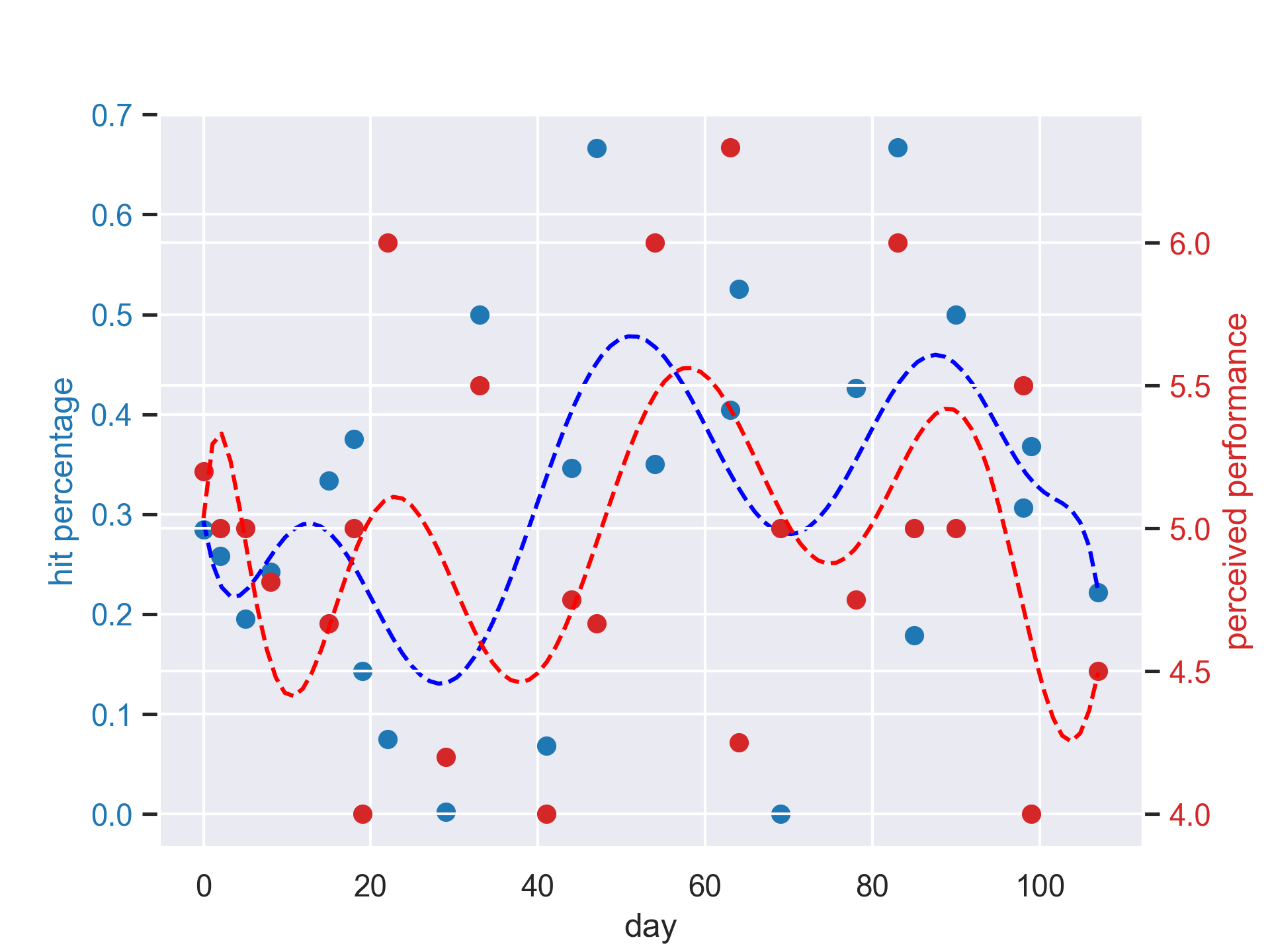}
        \caption{Hit scores (blue) and perceived performance reports (red). Daily team averages across the season timeline (Phase 4).}
        \label{fig:hits_season}
    \end{subfigure}%
    \hfill
    \begin{subfigure}[b]{0.48\columnwidth}
        \centering
        \includegraphics[width=\textwidth]{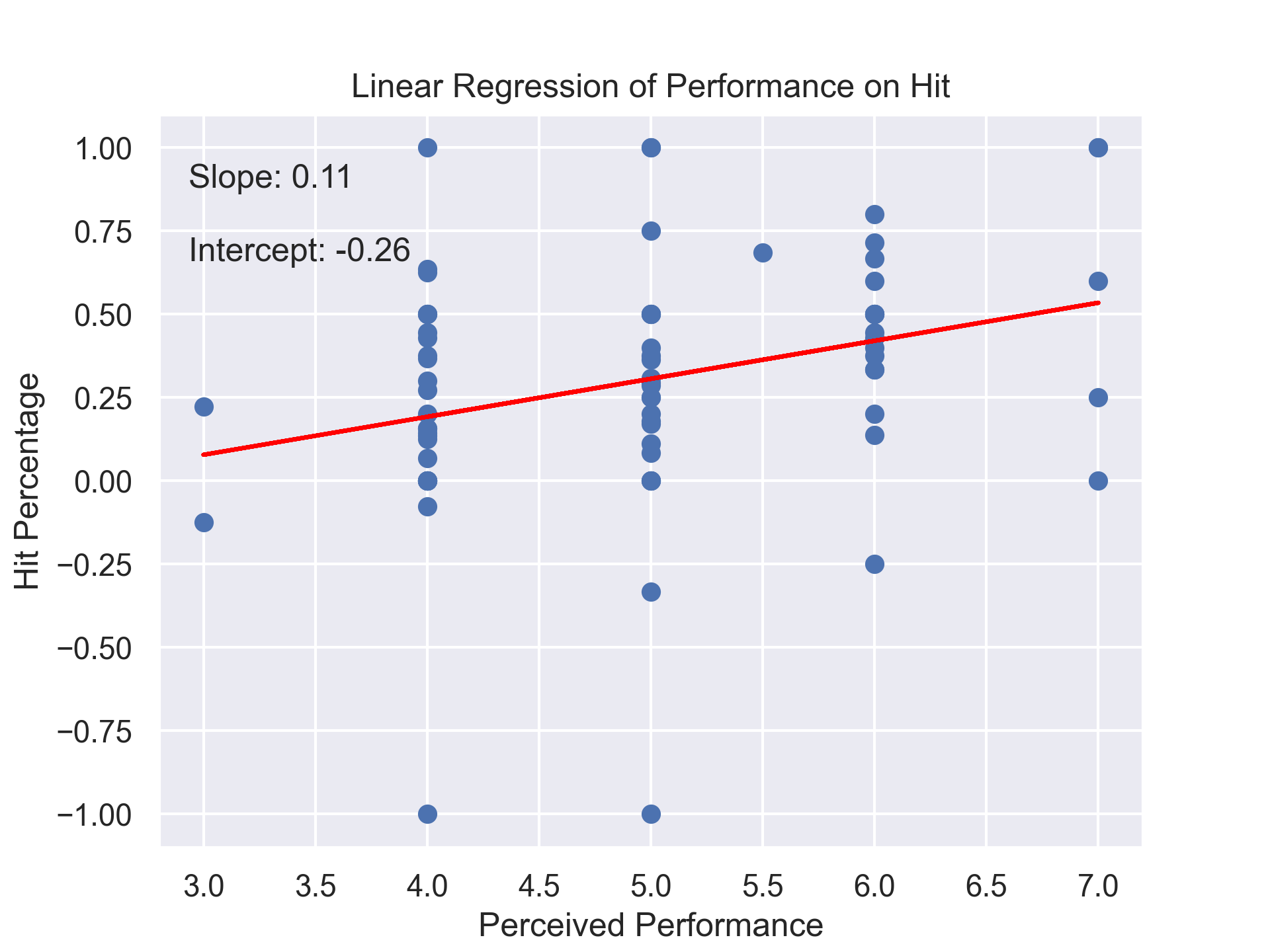}
        \caption{Linear regression fitted between the hit scores and perceived performance reports (slope = 0.11).}
        \label{fig:linear_reg}
    \end{subfigure}
    \caption{Linear Regression analysis of the relationship between hit percentage and perceived performance reports, individual data points (sample size: 76 perceived performance reports). }
    \label{fig:combined_plots}
\end{figure}

\textcolor{black}{In Figure \ref{fig:hits_season}, we present trends over the season (Phase 4) in a scatter plot (with fitted polynomials lines) to better observe the relationship between the hit percentage and reported perceived performance levels, across the team over the course of the season timeline. In Figure \ref{fig:linear_reg}, we present the linear regression analysis of perceived performance reports and the same day hit percentage scores from the matches. The fitted linear line has a slope of 0.11, which displays the existing relationship between the two variables.}  

\textcolor{black}{The analysis of the relationship between hit percentage scores and subjectively reported perceived performance levels through EMA reports during the season (Phase 4) reveals a strong positive correlation. This finding suggests that athletes' self-reported performance assessments align well with their actual performance, as measured by hit scores during matches. The results highlight the value of incorporating subjective performance measures alongside objective metrics to gain a more comprehensive understanding of an athlete's performance. Furthermore, the positive correlation between perceived soreness and hit scores underscores the importance of monitoring and managing physical stress and recovery.}

\subsection{Correlations between Hit Scores and Fitbit Features in Phase 4}

\textcolor{black}{The results show that several features related to oxygen saturation (SpO2) have significant negative correlations with hit scores. For example, \textit{SpO2 Std.} ($\rho$ = -0.430, $p$ = 0.006), \textit{SpO2-DFA-60} ($\rho$ = -0.424, $p$ = 0.006), and \textit{SpO2-DFA-40} ($\rho$ = -0.388, $p$ = 0.0135) show moderate to strong negative correlations. This suggests that players with higher variability or complexity in their SpO2 measurements tended to have lower hit scores.}

\textcolor{black}{Other sensor features that show significant positive correlations with hit scores include \textit{RHR change} ($\rho$ = 0.387, $p$ = 0.014), \textit{total sedentary time} ($\rho$ = 0.376, $p$ = 0.014), and \textit{SpO2 min} ($\rho$ = 0.350, $p$ = 0.027). These correlations indicate that players with larger changes in resting heart rate, more sedentary time, and higher minimum SpO2 values tended to have higher hit scores.}

\begin{table}[h]
\centering
\caption{Correlation Between Sensor Features and Hit Percentages Over the Season (Phase 4).}
\begin{tabular}{lrr}
\toprule
 & \textbf{$p$-value} & \textbf{Spearman's \( \rho \)} \\
\midrule
DFA-Sleep-10 & 0.047 & \colorcell{0.333} \\
SpO${_2}$-DFA-50 & 0.045 & \colorcell{-0.319} \\
SpO${_2}$-DFA-30 & 0.030 & \colorcell{-0.343} \\
SpO${_2}$-DFA-10 & 0.028 & \colorcell{-0.347} \\
SpO${_2}$ Min. & 0.027 & \colorcell{0.350} \\
Heart Rate Median & 0.026 & \colorcell{0.342} \\
Total Sedentary Time & 0.014 & \colorcell{0.376} \\
RHR Change & 0.014 & \colorcell{0.387} \\
SpO${_2}$-DFA-40 & 0.013 & \colorcell{-0.388} \\
Light Sleep Duration & 0.013 & \colorcell{-0.395} \\
SpO${_2}$-DFA-60 & 0.006 & \colorcell{-0.424} \\
SpO${_2}$ Std. & 0.006 & \colorcell{-0.430} \\
\bottomrule
\end{tabular}
\label{tab:fitbit_hit_phase4}
\end{table}

\textcolor{black}{Overall, these results suggest that various physiological and sleep-related factors, particularly those related to oxygen saturation and heart rate, are associated with volleyball hit scores.}

\section{Discussion}
\subsection{Wearable Devices as Predictive Tools for Volleyball Performance}
Our findings demonstrate that sensor data from wearable devices can predict season-long volleyball performance with generalizability to unseen participants. The 55-day pre-season period (Phase 2+3) yielded particularly strong predictive performance, achieving an F1 score of 0.75. Additionally, EMA data revealed that perceived productivity, recovery, and injury risk positively correlated with hit percentage, while perceived stress showed negative correlation.

\subsection{Evaluating the Benefits and Potential Drawbacks}
Our work advances upon previous volleyball performance prediction research \cite{de2022modeling} in several ways. While de Leeuw et al. used both sensor data from workouts and self-reported measures, our approach relies exclusively on passive wearable device data, enabling non-invasive continuous monitoring. We also implement LOSO for validation, enhancing generalizability to unseen participants.

Our results align with existing literature regarding sleep's impact on sports performance \cite{thun2015sleep, venter2012role} and the relationship between Heart Rate Variability (HRV), VO$_{2\text{max}}$, and athletic performance \cite{coyle1995integration, makivic2013heart}. However, counter-intuitive patterns in real-world data collection may be influenced by external factors such as individual skill levels, player positions, and academic commitments \cite{farrow2017development, lopes2020stress}.

Unlike previous models requiring historical performance data or human input \cite{claudino2019current, jung2021lax}, our approach enables immediate deployment for new participants, \textcolor{black}{helping coaches proactively identify players needing additional support. Nonetheless, implementation should carefully consider the user experience and psychological impacts.}

\subsection{Player Performance Trends}
\textcolor{black}{We also examined changes of player performance over time and by position, because prior work had shown activity patterns in sports can be position-specific \cite{wang2024toward}. Using Ordinary Least Squares (OLS) regression for time-series data on individual hit percentages, we found that although individual players showed varying trends, the average slope across the 14 players was not significantly different from zero ($p$ = 0.795), indicating no systematic, team-wide improvement or decline over the season. However, when analyzed by position, we have observed that middle hitters demonstrated statistically significant improvement over the course of the season \( (\beta = 0.0040, \rho=0.003) \), while other positions showed varying patterns: outside hitters maintained stable performance levels (effectively zero improvement due to a significant negative interaction term  \( \beta = - 0.0040, \rho = 0.013 \)), and setters showed a slight decline in performance (combined coefficient of -0.0015 per day). Our findings suggest that training and performance evaluation might benefit from position-specific approaches as well.}


\section{Limitations and Future Work}
Our study has several key limitations. The small sample size of 14 male athletes, while sufficient for preliminary exploration, limits result generalizability, suggesting the need for larger, more diverse cohort studies. Our reliance on Fitbit devices may not capture all relevant physiological and behavioral metrics, and the focus on automatically collected data excludes important variables like psychological factors and environmental conditions not quantifiable through wearable sensors.

While our modeling approach is rigorous, its applicability may be limited to volleyball-specific performance metrics. The study's external validity is constrained by the specific training environments and competitive contexts of collegiate volleyball. Additionally, unmeasured variables such as team dynamics and academic commitments, though potentially significant in athletic performance, are not accounted for in our current model. \textcolor{black}{Future work should plan to address these limitations by recruiting a broader and more diverse set of participants, spanning additional teams or leagues. It should also aim to incorporate richer contextual data (e.g., psychological states, academic load, position-specific training plans) to provide a more holistic view of each athlete. Future work can also explore training regression models to capture finer-grained performance levels, as well as experimenting with different neural network architectures with larger datasets (we observed that classical and tree based machine learning techniques worked much better than neural nets, in our dataset). Finally, it should seek to incorporate coach and athlete feedback loops, potentially integrating real-time monitoring and adaptive training regimens, to ensure that model outputs can be meaningfully interpreted and be put to use in day-to-day training.} 

\section{conclusion}
In conclusion, we employed machine learning models to predict the upcoming season's hit percentage average for individual players (good vs. poor hit average) using only passive sensing data collected from Fitbit wearable devices. Our best model, validated with LOSO, achieved promising predictive 
performance \textcolor{black}{(F1-score: 0.75)}. 
Next, we analyzed the relationships between self-reported psycho-physiological states and performance through correlation analyses. These findings hold the potential to guide the development of more customized coaching strategies and systems, thereby enhancing the effectiveness of training programs and \textcolor{black}{ advancing both the sports science and human activity and behavior computing communities. In real-world sports analytics, such a binary forecast could help coaches rapidly identify athletes at higher risk of under-performance and tailor targeted interventions or additional support. }


\bibliography{ref}
\end{document}